\documentclass[12pt]{article}

\title{Improved Inference for Respondent-Driven Sampling Data with Application to HIV Prevalence Estimation}

\author{Krista J.\ Gile\\Nuffield College\thanks{Krista J. Gile is Postdoctoral Prize Research Fellow, Nuffield College, Oxford.  The author would like to thank Mark S. Handcock of the University of Washington for numerous helpful suggestions, Tessie Caballero of COPRESIDA and Tetyana Saluk of AIDS Alliance for the use of their countries' datasets and Lisa G. Johnston of Tulane University for her help in suggesting, obtaining and assisting in the final cleaning of these datasets.  Direct correspondence to Krista J. Gile, Nuffield College, New Road, Oxford OX1 1NF, United Kingdom; e-mail krista.gile@nuffield.ox.ac.uk}}

\usepackage{soul}
\newcommand{\sout}{\st}

\usepackage{amssymb,amsmath,amsthm}
\usepackage{graphicx}
\usepackage{mathrsfs} 
\usepackage{natbib}
\usepackage{amsmath}
\usepackage{amssymb}
\usepackage[all]{xy}
\usepackage{color, graphics}
\usepackage{palatino, url, multicol}
\usepackage{graphicx}
\usepackage{subfigure}
\usepackage{mathrsfs}
\usepackage{multicol}
\bibpunct{(}{)}{;}{a}{,}{,}

\newcommand{\bi}{\begin{itemize}}
\newcommand{\ei}{\end{itemize}}
\definecolor{Emphcolor}{cmyk}{0,0.89,0.94,0.1}
\definecolor{Netcolor}{rgb}{.8,0,.9}
\definecolor{Diseasecolor}{rgb}{1,.8,.2}
\definecolor{Sampcolor}{rgb}{0,.9,.3}
\definecolor{Black}{rgb}{0,0,0}
\definecolor{Red}{rgb}{1,0,0}
\definecolor{Blue}{rgb}{0,0,1}
\definecolor{Gray}{gray}{.6}

\newcommand{\gray}{\color{Gray}}

\newcommand{\IR}{\mathbb{R}}

\font\elevenrm=cmr11
\newcommand{\ql}{{\elevenrm ``}}
\newcommand{\qr}{{\elevenrm "}\ }

\theoremstyle{plain}

\newcommand{\N}{\mathbb{N}}

\theoremstyle{definition}

\graphicspath{{../figures/}}

\usepackage{graphicx}
\usepackage{float}
\usepackage{rotating}
\usepackage{natbib}
\usepackage{color}
\usepackage{hypernat}
\usepackage{pdflscape}

\usepackage{setspace}
\usepackage{graphics}
\usepackage{subfigure}
\usepackage[all]{xy}
\usepackage{amsmath}
\usepackage{amssymb}
\usepackage[all]{xy}
\usepackage{palatino, url, multicol}
\usepackage{mathrsfs}
\usepackage{multicol}
\usepackage{enumerate}

\definecolor{Emphcolor}{rgb}{.1,.1,.5}
\definecolor{Red}{rgb}{.9,0,.1}
\definecolor{Blue}{rgb}{.1,.1,.5}

\newcommand{\vg}{{\bf g}}

\newcommand{\vz}{{\bf z}}

\newcommand{\bea}{\begin{eqnarray}}
\newcommand{\eea}{\end{eqnarray}}

\graphicspath{{../figures/}}

\usepackage{ifthen}
\newboolean{draft}
\setboolean{draft}{false}
\newcommand{\knote}[1]{\ifthenelse{\boolean{draft}}{{\bf knote:~}{\it
#1}\relax}{}}

\newboolean{blind}
\setboolean{blind}{false}
\newcommand{\blinded}[1]{\ifthenelse{\boolean{blind}}{}{#1}}
\newcommand{\unblinded}[1]{\ifthenelse{\boolean{blind}}{#1}{}}

\newcommand{\mvh}{\hat{\mu}_{\rm VH}}

\newcommand{\mh}{\hat{\mu}_{SS}}

\newcommand{\vp}{{\bf p}}
\newcommand{\vG}{{\bf G}}
\newcommand{\vv}{{\bf v}}
\newcommand{\vV}{{\bf V}}
\newcommand{\vU}{{\bf U}}
\newcommand{\vS}{{\bf S}}
\newcommand{\vd}{{\bf d}}
\newcommand{\vpi}{{\mbox{\boldmath$\pi$}}}

\begin{document}

\maketitle


\vspace{-.25in}
\begin{abstract}
Respondent-driven sampling is a form of link-tracing network sampling, which is widely used to study hard-to-reach populations, often to estimate population proportions.  Previous treatments of this process have used a with-replacement approximation, which we show induces bias in estimates for large sample fractions and differential network connectedness by characteristic of interest.

We present a treatment of respondent-driven sampling as a successive sampling process. Unlike existing representations, our approach respects the essential without-replacement feature of the process, while converging to an existing with-replacement representation for small sample fractions, and to the sample mean for a full-population sample.

We present a successive-sampling based estimator for population means based on respondent-driven sampling data, and demonstrate its superior performance when the size of the hidden population is known.  We present sensitivity analyses for unknown population sizes.  In addition, we note that like other existing estimators, our new estimator is subject to bias induced by the selection of the initial sample.  Using data collected among three populations in two countries, we illustrate the application of this approach to populations with varying characteristics.  We conclude that the successive sampling estimator improves on existing estimators, and can also be used as a diagnostic tool when population size is not known.

\end{abstract}



\doublespacing

\section{Background}
Researchers and government officials are often interested in characteristics of human populations for which there are no practicable sampling frames for direct sampling.  In some such hidden populations, members are connected through social networks.  A common approach is to collect a sample using a variant of {\it link-tracing} \citep{Thompson2006, Thompson2006a}, such as a {\it snowball sample} \citep{goodman1961}, where subsequent sample members are selected based on their relations with previously sampled individuals.  When the initial sample is not a probability sample, this approach does not result in a probability sample.  However, most available alternatives, \citep{muhib01,peterson_etal2008, targetedsampling1989} also fail to produce a probability sample of the population.

Respondent-Driven Sampling (RDS, introduced by Heckathorn 1997, 2002, see also Salganik and Heckathorn, 2004, Volz and Heckathorn 2008\nocite{heck97, Heckathorn2002, salgheck04, volzheck08}) is a recently introduced variant of link-tracing sampling which increases the ease of sampling and 
produces samples which, it is argued, approach probability samples as sampling progresses.
The lack of satisfactory alternatives has spurned a strong demand for RDS \citep{johnston08, lanskycdcrds07}.

RDS presents two main innovations:
a sampling design and a corresponding approach to estimation.  The former is highly effective in many settings \citep{lanskycdcrds07};  the {\it respondent-driven} design relies on the respondents at each wave to select the next wave by distributing uniquely identified coupons to others in the target population, who can choose to return the coupons to enroll in the study.  Thus, the sampling exploits the network of social relations while also avoiding the confidentiality concerns associated with recording the names of contacts.  

The key innovation for estimation is that through many waves of sampling, the dependence of the final sample on the initial sample is reduced, allowing researchers more confidence in making approximate probability statements about the resulting samples.  This insight allows for statistical inference in settings where the initial sample is typically selected by a convenience mechanism.  Although current inference is likely superior to the alternative non-probability methods, existing methods are sensitive to deviations from many assumptions \citep{goel2007, gilehanSM09, neely09}.\knote{after blind:  add Gile 2008 reference}  This paper offers a modification of the existing theoretical formulation of respondent-driven sampling, and corresponding inference
to address what we find to be a serious conceptual weakness of existing work: the known inaccuracy of the with-replacement approximation to the sampling process.

In the next section, we begin by introducing current RDS estimation, particularly the estimator introduced by \cite{volzheck08}, and illustrate the sensitivity of this estimator to the with-replacement sampling assumption in cases of substantial sample fractions.  In Sections \ref{sssection} and \ref{ssestsection}, we then introduce a new model for RDS sampling based on successive sampling,
and introduce a new estimator based on that model.  In Section \ref{mhsim}, we use a simulation study to illustrate the superior performance of the new estimator.  We also include sensitivity analyses concerning inaccurate estimation of the size of the target population and the characteristics of the initial sample.  In Section \ref{sec:apply}, we apply our estimator to data collected in three populations of drug users and men who have sex with men.  We conclude with a broader discussion of the method and its limitations.

\section{Previous Approaches to Estimation} \label{sec:rdsest}
The basic ideas underlying estimation from RDS data are clever and important. They 
allow for something like valid statistical inference, in a sampling setting where the target population cannot be effectively reached using a traditional sampling frame.  
The original article, \citep{heck97} made very strong assumptions about the sampling procedure so as to assume that the sample proportions were representative of the population proportions.  \cite{salgheck04} introduced a Markov chain argument for population mixing, and proposed an estimator based on equating the number of cross-relations between pairs of sub-populations of interest, based on the referral patterns of each group.  This estimator is currently in wide use, and is implemented in the standard RDS analysis software \citep{rdsat}.  \cite{volzheck08} connect RDS estimation to mainstream survey sampling through the use of a generalized Horvitz-Thompson estimator form.  This estimator relies on the estimation of the inclusion probabilities of the sampled units, $\pi_i$.  Based on an argument for treating the sample as independent draws from the stationary distribution of a random walk on the nodes of an undirected graph, \cite{volzheck08} approximate the sampling probabilities as proportional to nodal degree, $d_i$, or number of incident edges in the graph.  This estimator avoids the problem of potentially unknown population size $N$ by using the generalized Horvitz-Thompson form, normalizing by an estimator of $N$ as follows:
\bea
\hat{\mu}_{\rm VH} = \frac{\sum_{i=1}^N\frac{\vS_i \vz_i}{\vd_i}}{\sum_{i=1}^N\frac{\vS_i}{\vd_i}},
\label{volzheckest}
\eea
where $\vS_i=1$ indicates that the $i^{th}$ unit has been selected for sampling, and $\vS_i=0$ indicates it has not been selected, and $\vz_i$ represents the variable of interest measured on the $i^{th}$ unit.

We refer to this approach as the Volz-Heckathorn (VH) estimator.  \cite{gilehanSM09} illustrate that this estimator consistently out-performs the Salganik-Heckathorn estimator, and we believe it is the most principled estimator currently available for RDS.
For these reasons, we use the VH estimator as the standard for comparison in this paper.

Many critical assumptions required to justify this estimator are explored in \cite{gilehanSM09} and listed in Table \ref{tab:assmh}.  In this paper, we focus on the reliance on a with-replacement sampling model.  We present an estimator based on an alternative sampling model reflecting the without-replacement nature of the sampling process.

\section{Successive Sampling for RDS}\label{sssection}

The Volz-Heckathorn estimator requires many waves of sampling to justify its reliance on a stationary distribution.  In practice the number of waves is small (almost always fewer than 20, and often 5 or fewer).  Also, we wish to consider a without-replacement process for which stationarity does not apply. It is therefore instructive to consider a special case when stationarity is not necessary.

Consider a graph formed in the following manner:  Begin with $N$ vertices, designated by indices $1:N$.  Assign to each vertex a number of edge-ends according to arbitrary fixed degree distribution $\N=\N_1, \N_2, \ldots \N_K$, where $\N_j$ is the number of nodes of degree $j$, $1$ and $K$ are the minimum and maximum degrees, respectively, and subject to the constraint that twice the number of edges, $2E=\sum_{k \in 1 \ldots K}k \N_k$, is even.  Now select pairs of edge-ends completely at random and assign an edge connecting each pair.  This procedure results in a variant of the so-called {\it configuration model} for networks, a popular null model for networks, especially in the physics literature \citep{molloyreed1995}.  Note that this formulation does allow {\it loops}, or links to oneself as well as multiple edges between the same pair of vertices, although the rate of these events decreases  with increasing population size for fixed maximum degree $K$, such that several authors have suggested they are negligible for $K < (\bar{\vd}N)^{1/2}$ or $K < N^{1/2}$, where $\bar{\vd}$ is the mean degree of the network \citep{chunglu2002, burdakrzy2003, bogunapv2004, catanzarobp2005, fosterfgp2007}.

Now consider a random walk $\vG^*$ on a set of vertices with degrees given by $\vd_1, \vd_2, \ldots, \vd_N$, such that $\sum_i \mathbb{I}(\vd_i=k)=\N_k$, where $\mathbb{I}(A)$ is the indicator function on $A$.  \cite{volzheck08} consider the stationary distribution of this random walk for a fixed graph.  Instead, we consider the transition probabilities of the corresponding walk over the distribution of all networks of fixed degree distribution constructed as above, in which the $j^{th}$ node visited, $\vG^*_j$, is selected from the distribution of possible edges from node $\vG^*_{j-1}$.  The transition probabilities are then given by:
\bea
P(\vG^*_j=\vg^*_j | \vG^*_1,\vG^*_2, \ldots \vG^*_{j-1} = \vg^*_1,\vg^*_2, \ldots \vg^*_{j-1}) = \left\{
\begin{array}{cl}
\frac{\vd_{\vg^*_j}}{2E-1} & \vg^*_j \neq \vg^*_{j-1} \\\\
 \frac{\vd_{\vg^*_j}-1}{2E-1} & \vg^*_j = \vg^*_{j-1},
\end{array}
\right.
\eea
where this probability is taken over the space of all possible configuration model graphs of given degree distribution, as well as over the steps of the random walk.

This procedure results in stationary distribution proportional to $\vd_i$, and, in fact, selection probabilities at each step very nearly proportional to $\vd_i$.  
Thus, for a network structure given by such a configuration model, the Volz-Heckathorn estimator constitutes a \cite{HansenHurwitz1943} estimator, without further requirement of sufficient waves for convergence. 

Now consider the corresponding self-avoiding random walk $\vG$.  In this procedure,
\bea
P(\vG_j=\vg_j | \vG_1, \vG_2, \ldots \vG_{j-1} = \vg_1, \vg_2, \ldots \vg_{j-1}) = \left\{
\begin{array}{cl}
\frac{\vd_{\vg_j}}{2E-\sum_{i = 1}^{j-1}\vd_{\vg_i}} & \vg_j \notin \vg_1 \ldots \vg_{j-1} \\
0 & \vg_j \in \vg_1 \ldots \vg_{j-1}.
\end{array}
\right. \label{ppsworeqn}
\eea
This sampling procedure is mathematically equivalent to {\it successive sampling} (SS) or {\it probability proportional to size without replacement sampling} (PPSWOR), dating back to \cite{yatesgrundy53}, and typically defined by the following sampling process:

\bi
\item Begin with a population of $N$ units, denoted by indices $1\ldots N$ with varying sizes represented by $\vd_1, \vd_2, \ldots \vd_N$, with $\sum_{i=1}^N \vd_i = 2E$, for total edges $E$.
\item Sample the first unit $\vG_1$ from the full population $\{1 \ldots N\}$ with probability proportional to size $\vd_i$.
\item Select each subsequent unit with probability proportional to size {\it from among the remaining units}, such that conditional sampling probabilities are given by (\ref{ppsworeqn}).
\ei

In the survey sampling literature, mostly based on the work of \cite{raj56} and \cite{murthy57}, this sampling design is referred to as {\it probability proportional to size without replacement} (PPSWOR), and typically used in instances where the desired {\it probability proportional to size} (PPS) design is not feasible.  In such cases, the sizes of population units are all known, the sampling design is implemented, and the main interest is in estimating the population total or mean of a variable measured on sampled units.  The analytical properties of this procedure are quite difficult, as suggested by more recent work including \cite{raosensin91} and \cite{kockor01}. In fact, even the marginal unit sampling probabilities are not available in closed form. \knote{insert the other one mark found more recently?}

In the geological discovery literature, successive sampling is not a purposive design, but an approximation to a non-designed sampling process.  Important work in this area includes \cite{andkau86}, \cite{nairwang89}, and \cite{bnw92}.  These authors address the case of oil field discovery, in which successive fields are discovered with probabilities proportional to some measure of their size, typically taken to be their volume. This literature does not assume the full population of sizes to be known, and typically takes some function of the sizes, such as the sum of the sizes of undiscovered reserves, as the object of inference.  Our application of successive sampling also assumes the unobserved sizes to be unknown, however as in \cite{raj56} and \cite{murthy57}, our object is to use these sizes to estimate the population characteristics on another variable measured on all sampled units.  To do so, we both develop a novel algorithm and leverage more recent work by \cite{fattorini06}.

\section{Estimation of population means from RDS samples based on successive sampling}
\label{ssestsection}
\newcommand{\f}{f_{\vpi}(k;n,\N)}

Under successive sampling, for population of sizes given by $\N$ and sample size $n$, there is a function $\f$ mapping the size $k$ of a unit to its sampling probability $\vpi_i$.  Our proposed estimator is based on estimated sampling weights, which are based on the $\vpi$ given by the successive sampling procedure, applied to nodal sampling units and {\ql}sizes\qr given by nodal degrees.  There are three key challenges for this approach.  The mapping depends on first, the known population size $N$ and, second, the degree distribution, $\N$, neither of which is known in the general RDS case.  Finally, given $\N$ and the sample size $n$, the mapping is not explicit.
The lack of explicit mapping has been addressed by \cite{fattorini06}, who suggests estimating the mapping by simulation.  For known $\N$ and given $n$, he simulates the successive sampling procedure, then estimates the inclusion probability $\vpi_i$ associated with unit $i$ by:
\bea
\tilde{\vpi}_i = \frac{\vU_i + 1}{M+1},
\label{fattorinip}
\eea
where $\vU_i$ is the number of times unit $i$ is sampled in the $M$ trials.  He proposes using these estimated probabilities in the standard Horvitz-Thompson estimator:
\bea
\sum_{j: \vS_j = 1}\frac{\vz_j}{\tilde{\vpi}_j}.
\label{fattoriniT}
\eea

\knote{should we have different notation for the sampling probabilities under different procedures?}

\newcommand{\E}{\mathbb{E}}
\newcommand{\hatqdn}{{\hat{\pi}}}
\def\E{\mathop{\rm E\,}\nolimits}

In most of this paper, we assume the population size, $N$, is known.  We evaluate the sensitivity of our results to that assumption in Section \ref{sec:sizesens}, and in the examples in Section \ref{sec:apply}.  

Given the known population size, we present a novel approach to estimating the degree distribution $\N$ jointly with inclusion probabilities when degrees are only observed for sampled nodes.  This procedure can be applied beyond the RDS setting whenever the population distribution of unit sizes corresponding to a sample collected through successive sampling is unknown.

\newcommand{\q}{\f}
\newcommand{\qnk}{f_{\vpi}(k;n,\N)}

Our approach iteratively estimates the population distribution $\N$ and the mapping $\qnk:k \to \vpi$.
This approach relies on two key points:
\bi
  \item For known population of degrees $\N$, the mapping $\q$ can be estimated by simulation in a form similar to (\ref{fattorinip}).
  \item For known mapping $\q$, the number (or proportion) of the population of degree $k$ can be estimated using a form similar to (\ref{fattoriniT}).
\ei

We leverage these two points to propose the following procedure for the estimation of population mean $\mu$ in the case of nodal degrees observed only for sampled units.
\knote{all this notation needs to be improved}

Let $\E[\cdot ; n, \N]$ denote expectation with respect to a sample of size $n$ sampled by successive sampling from a population with degree counts $\N=\{\N_1, \N_2, \ldots, \N_K\}.$
Then:
\bea
\E[\vV_{k} ; n, \N] = \N_{k} \qnk
~~~~~~~~~k=1, \ldots, K
\eea
where $\vV_k$ is the random variable representing the number of sample units with degree $k$,  
$\qnk=\E[\vS_{j}:\vd_{j}=k; n, \N]$ is the (common) inclusion probability of a
node of degree $k$, and $K$ is the maximum degree, $K<N$.
This suggests first order moment equations for the unknown true $\N:$
\bea
\E[\vV_{k} ; n, \N] = \vv_{k}
~~~~~~~~~k=1, \ldots, K
\label{mmeqn}
\eea
where $\vv_k$ is the observed number of sample units with degree $k$.

\newcommand{\qnko}{f_{\vpi}(k;n,\N^0)}
\newcommand{\qnki}{f_{\vpi}(k;n,\N^i)}
\newcommand{\qnkim}{f_{\vpi}(k;n,\N^{i-1})}
\newcommand{\qnlim}{f_{\vpi}(l;n,\N^{i-1})}
\newcommand{\qndotr}{f_{\vpi}(\cdot;n,\N^r)}

The algorithm is then:
\begin{enumerate}
\item Initial estimate:
\bea
\qnko = \frac{k}{N} \sum_{l =1}^K \frac{v_l}{l},
\eea
that is $\qnko$ proportional to $k$.
\item For $i=1, \ldots, r$ iterate the following steps:
\begin{enumerate}
  \item Estimate the population distribution of degrees:
\bea
{\N_{k}^{i} = N \cdot \frac{\frac{\vv_k}{\qnkim}}{\sum_{l =1}^K \frac{\vv_l}{\qnlim}}  }~~~~~~~~~~~k=1, \ldots, K
\eea
{where $\vv_k$ is the observed number of sample units with degree $k$.}
  \item Estimate inclusion probabilities:
\begin{enumerate}
  \item Simulate $M$ successive sampling samples of size $n$ from a population with
        composition  $\N^{i}.$
  \item Estimate the inclusion probabilities:
\bea
\qnki = \frac{\E[\vV_{k} ; n, \N^{i}]}{\N_{k}^{i}} \approx \frac{\vU_k + 1}{M \cdot \N_{k}^{i} + 1}, \label{estq2a}
\eea
{\small where $\vU_k$ is the total number of observed units of size $k$ in the $M$ simulations.}
\end{enumerate}
\end{enumerate}
\item Estimate the population distribution of degrees and the corresponding
inclusion probabilities via, respectively:
${\hat{\N}} = \N^r$ and $\hatqdn(\cdot) = \qndotr.$
\item Use the resulting mapping $\hatqdn$ to estimate $\mu$ via the
generalized Horvitz-Thompson estimator:
\bea
\mh = \frac{\sum_{j = 1}^N \frac{\vS_j \vz_j}{\hatqdn(\vd_j)}}{\sum_{j= 1}^N\frac{\vS_j}{\hatqdn(\vd_j)}}.
\label{estmu2}
\eea
\end{enumerate}

For computational efficiency, most simulations in this paper were conducted with $M=500$ and $r=3$, with good results.
In general, we recommend at least $M=2000$ and $r=3$, and we have used these parameters for the simulations of the standard error procedure in the supplemental materials, in the application to real data, and in the extension to $N>1000$ in the discussion.  Estimation time scales with sample size, population size, and $M$.  In our simulations, with $N=1000$ and $M=500$, estimates require about $1.5$ seconds on a personal computer, increasing to about $6$ seconds when $M=2000$.  In practice, these parameters can be adjusted for desired precision in the solution to (\ref{mmeqn}).  Higher values of $M$ are particularly helpful for more dispersed degree distributions and larger population sizes.    

Simulations have shown the results of this procedure to be at least as good as those provided by a first order asymptotic approximation provided by \cite{andkau86}.  This approach is novel and its theoretical properties are not well understood.  
It is appealing is to consider the algorithm as a variant of an EM algorithm, such as an ECM algorithm \citep{ECM}, with the unobserved part of the degree sequence, $d_{n+1}, \ldots d_N$ as the latent variable.  Unfortunately, in the design-based frame, these values also fully determine the unknown parameters $N_k = \sum_{i=1}^N {\mathbb I}(d_i=k)$, resulting in a degenerate likelihood form.

\subsection{Between Infinite Population and Full Population Sample}

Consider the limiting case of the moment equation (\ref{mmeqn}) where $N=n$.  In this case, this equation is only satisfied for the degree distribution given by, $\hat{\N}_k = \vv_k$, resulting in $\hatqdn(\vd_j) = 1 ~\forall ~ j$, and $\mh = \hat{\mu}$, the sample mean.

Now consider the limit as $N \to \infty$, for fixed $n$ and fixed maximum degree.  
Then the step-wise selection probabilities for an unsampled node approach values proportional to degree:
\bea
\frac{\vd_i}{\sum_j \vd_j - \sum_{j: \vS_j=1} \vd_j} - \vp_i \to 0,
\eea
where $\vp_i = \frac{\vd_i}{\sum_j \vd_j}$.
Then $P(\rm{Binom}(n, \vp_i)>1) \to 0$, such that
\bea
\frac{\vpi_i}{\vpi_j} \to \frac{\vp_i}{\vp_j} = \frac{\vd_i}{\vd_j},
\eea
for overall inclusion probabilities $\vpi$, and step-wise selection probabilities $\vp$.
Therefore, $\mh \to \mvh$.

In either limit, $N \to n$ or $N \to \infty$, $\mh$ approaches an existing estimator.  Thus, it retains the professed limiting properties of $\mvh$, such as robustness to bias based on the initial sample, while retaining the favorable finite population characteristics of the sample mean in the case of a large sample fraction.  In the next section, we use simulation studies to argue that for $n < N < \infty$, the proposed estimator appropriately mediates between these two, and therefore out-performs both.

\knote{se was here}

\section{Comparing the New and Existing Estimators:\\ A Simulation Study} \label{mhsim}
Our simulation study is designed to highlight the treatment of without-replacement sampling in the case of a large sample fraction.
To increase the realism of the study, we chose parameters to match the characteristics of the pilot data from the CDC surveillance program \citep{aqcdc06} wherever possible.  The general procedure was as follows:
\bi
\item 1000 networks are simulated under each test condition.
\item An RDS sample is simulated from each sampled network.
\item RDS estimators are computed from each sample.
\ei

Because the CDC's surveillance system aims for a sample size of 500, and many RDS studies approach exhaustion of their populations of interest \citep{johnston09exhaust, cdc08exhaust}, we fix all sample sizes at 500.  We also consider a mean degree of 7, close to the mean of the pilot data from the CDC study \citep{aqcdc06}.

We assign a discoverable class to each member of the simulated population.  In reference to studies designed to estimate the prevalence of infectious disease, we refer to this characteristic as {\ql}infection status,\qr  assigning the {\ql}infected\qr status ($\vz_i=1$) to 20\% of simulated population members in each simulation.  Note that $\mh$ could also be applied to a continuous or categorical variable.

\newcommand{\y}{{\bf y}}
\newcommand{\Y}{{\bf Y}}
\newcommand{\x}{{\bf x}}
\newcommand{\veta}{{\mbox{\boldmath$\eta$}}}
\newcommand{\g}{{\bf g}}
\newcommand{\uu}{{\bf u}}

We consider networks sampled from models from the {\it Exponential-family Random Graph Model}
(ERGM) class \citep{sprh06}. Here the relations $\y$ are represented as a realization
of the random variable $\Y$ with distribution:
\begin{eqnarray}
P_{\eta}(\Y=\y | \x) = \exp\{\veta{\cdot}\g(\y,\x)-\kappa(\veta,\x)\}\quad \quad \y\in {\cal
Y},
\label{ergm}
\end{eqnarray}
where $\x$ are nodal or dyadic covariates, $\g(\y,\x)$ is a $p$-vector of network statistics,
$\veta\in \IR^p$
is the parameter vector, ${\cal Y}$ is the
set of all possible undirected graphs, and
$\exp\{\kappa(\veta,\x)\} =
\sum_{\uu\in{\cal Y}}\exp\{\veta{\cdot}\g(\uu,\x)\}
$
is the normalizing constant \citep{bar78}.
The structure of the networks represented is determined by the choice of
$\g(\y,\x)$.

In this study, we vary the structure of the networks in three ways:  First, we consider populations of sizes (i.e. numbers of nodes) 1000, 835, 715, 625, 555, and 525, such that samples of size 500
constitute about 50\%, 60\%, 70\%, 80\%, 90\%, and 95\% of the target population.  We focus on this range of sample fractions because it highlights settings in which we might expect to see finite population biases such as those $\mh$ is intended to address.  Another critical feature is the {\it activity ratio} $w$, equal to the ratio of the mean degree of infected nodes to the mean degree of uninfected nodes, 
\bea
w=\frac{\sum_{i=1}^N d_i z_i }{\sum_{i=1}^N z_i }
\frac{\sum_{i=1}^N (1-z_i)}{\sum_{i=1}^N d_i (1-z_i)}.
\eea
This measures the differential tendency for the groups to be socially connected in the population.  We consider $w \in \{0.5, 0.8, 1, 1.1, 1.4, 1.8, 2.5, 3\}$. 

We also induce homophily on infection status in these simulations, parameterized as the relative probability of an edge between two infected nodes, and an edge between an infected and an uninfected node.  This is an intuitive parameterization, also used in \cite{gilehanSM09}, but differs from that used in other analyses of RDS.  Except in Section \ref{sec:senshomoph},  the edge probability between the two infected nodes is fixed at five times that of the mixed dyad.  
For $w=1$, this, along with the 20\% infected,  implies that an edge between two uninfected nodes is twice as likely as an edge in a mixed dyad.  We consider several other levels of homophily, and in Section \ref{sec:senshomoph} we show that this feature has important implications for the bias induced by biased initial samples, however for the case of initial samples at random with respect to $\vz$, as in most of the simulations presented here, bias was not affected by level of homophily, although increasing homophily does increase the variance of both $\mh$ and $\mvh$, and is therefore important to consider in standard error estimation. \knote{could show a variance simulation for the change in variance estimation (width and coverage) for changing homophily.}

These features are represented in the ERGM by choosing
network statistics to represent the mean degree, the activity ratio $w$, and homophily (based on
using the {\ql}infected{\qr} status as a nodal covariate).  These three values were specified using the {\ql}{\tt nodemix}\qr {\tt statnet} model term, which includes three parameters corresponding to the three cells of the mixing matrix on infection.
 The parameter $\veta$ was chosen
so the expected values of the statistics were equal to the values given
above \citep{vanduijngilehan09}.  Samples from the resulting models were taken using the {\tt statnet} R package \citep{statnet,ergmjss}.

The RDS sampling mechanism is again designed to mimic that of the CDC's pilot study.   Ten initial sample nodes were chosen for each sample, selected sequentially 
with probability proportional to degree, (i.e. by successive sampling).
In the sensitivity analysis, we also consider initial sample selection regimes dependent on $\vz$.
Subsequent sample waves were selected without-replacement by sampling up to two nodes at random from among the un-sampled alters of each sampled node.  Exactly two alters were sampled whenever possible.  This process typically resulted in the sampling of four complete waves and part of a fifth wave, stopping when a sample size of 500 was attained.

We augment our basic results with two sub-studies evaluating the sensitivity of the estimator $\mh$ to assumptions not required for $\mvh$:  the accuracy of the assumed population size $\hat{N}$ and the dependence on the initial sample.

\subsection{Results}\label{sec:basic}
\begin{figure}[h]
\begin{center}
    \includegraphics[width=3in]{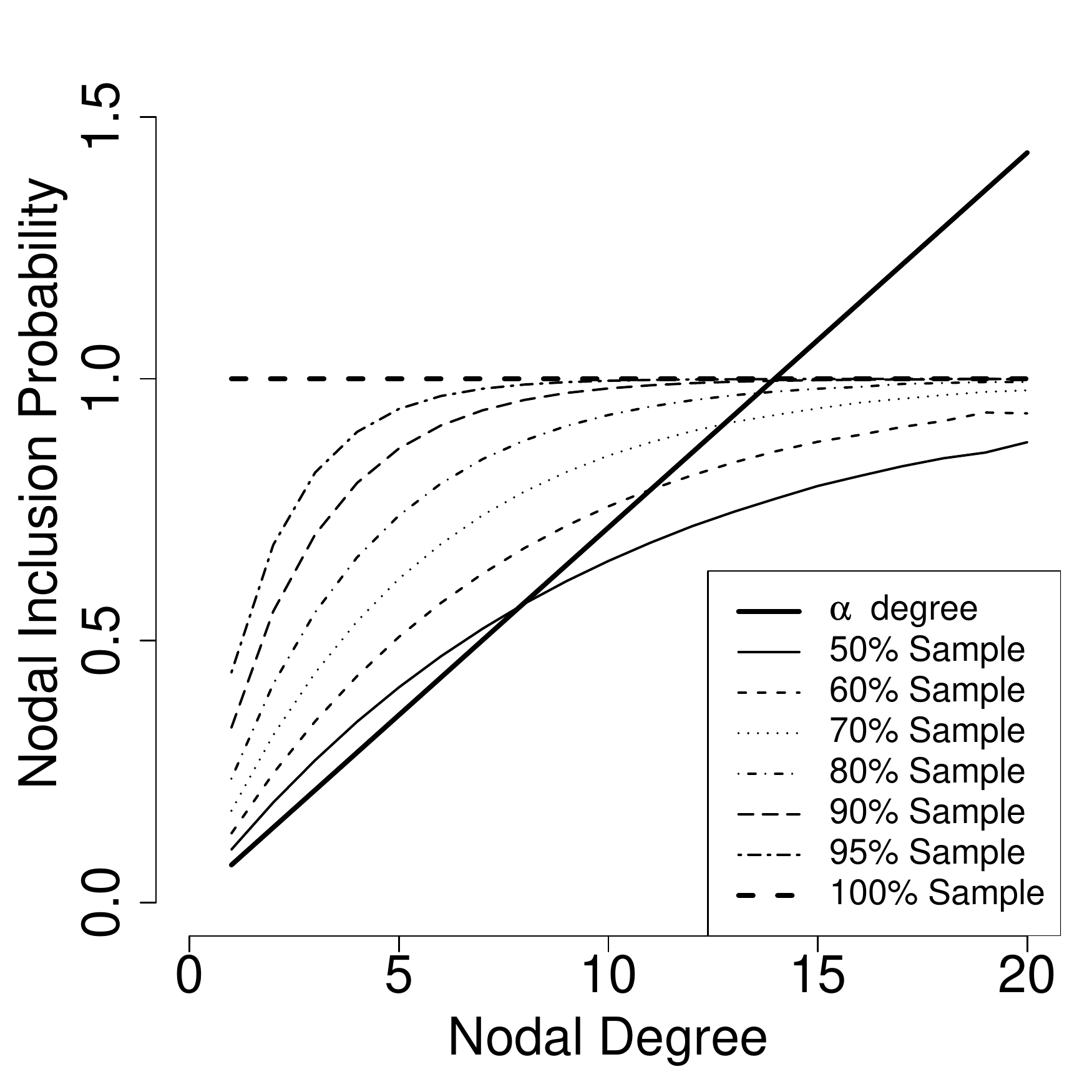}
\end{center}
\caption{Estimated mappings from nodal degree to inclusion probability for successive sampling samples of size 500 constituting about  50\%, 60\%, 70\%, 80\%, 90\%, 95\%, and 100\% of the population, for simulated network degree distributions with $w=1$, along with the proportional mapping assumed by the Volz-Heckathorn estimator (for 50\% sample), indicated by {\it $\propto$ degree}.  Note that given the simulated degree distribution, the proportional mapping requires probabilities greater than 1 to attain the desired sample size.} \label{fig:curvesboth}
\end{figure}

The Volz-Heckathorn estimator exhibits substantial bias in cases of non-unity activity ratio, and more so for larger sample fractions.  This result was noted in \cite{gilehanSM09}, and is illustrated in Figure \ref{fig:biasbars}.  The bias can be understood as follows:  in the case of higher mean degree among infected nodes $(w>1)$, and large sample fraction, the higher-degree infected samples will be down-weighted proportional to their degrees.  The true without-replacement sampling probabilities are closer to uniform than would be suggested by the proportional-to-degree estimates, such that these higher-degree nodes are excessively down-weighted, leading to negative bias in the estimated proportion infected.  The corresponding mappings from degree to sampling weight are illustrated in Figure \ref{fig:curvesboth}.

\newcommand{\kangaroo}{7.2cm}

\begin{figure}[h]
\begin{center}
\subfigure[Bias of $\hat{\mu}_{VH}$]
{
    \label{fig:biasbars}
    \includegraphics[width=\kangaroo]{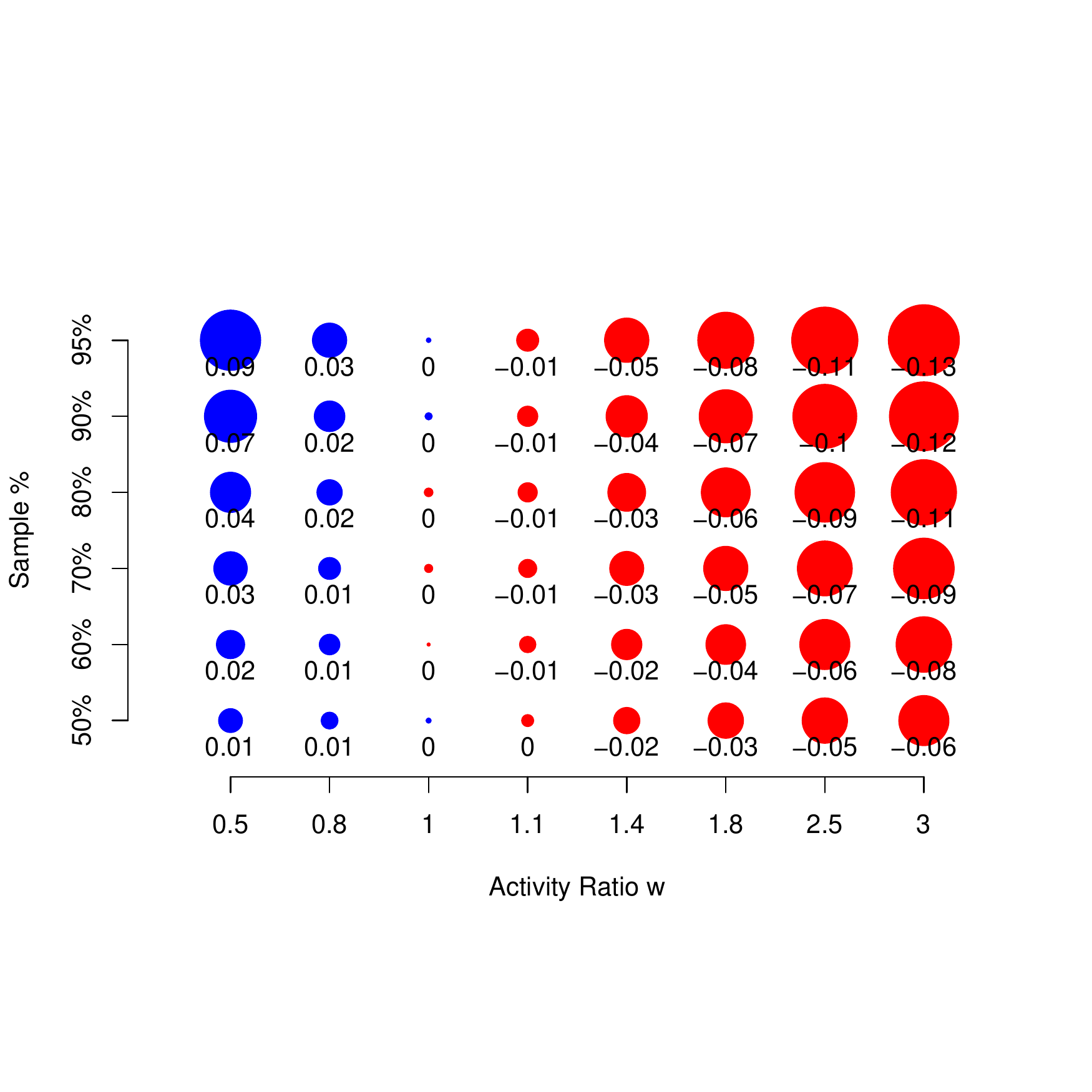}
} \hspace{-.3cm}
\subfigure[Bias of $\mh$]
{
    \label{fig:sppsbiasbars}
    \includegraphics[width=\kangaroo]{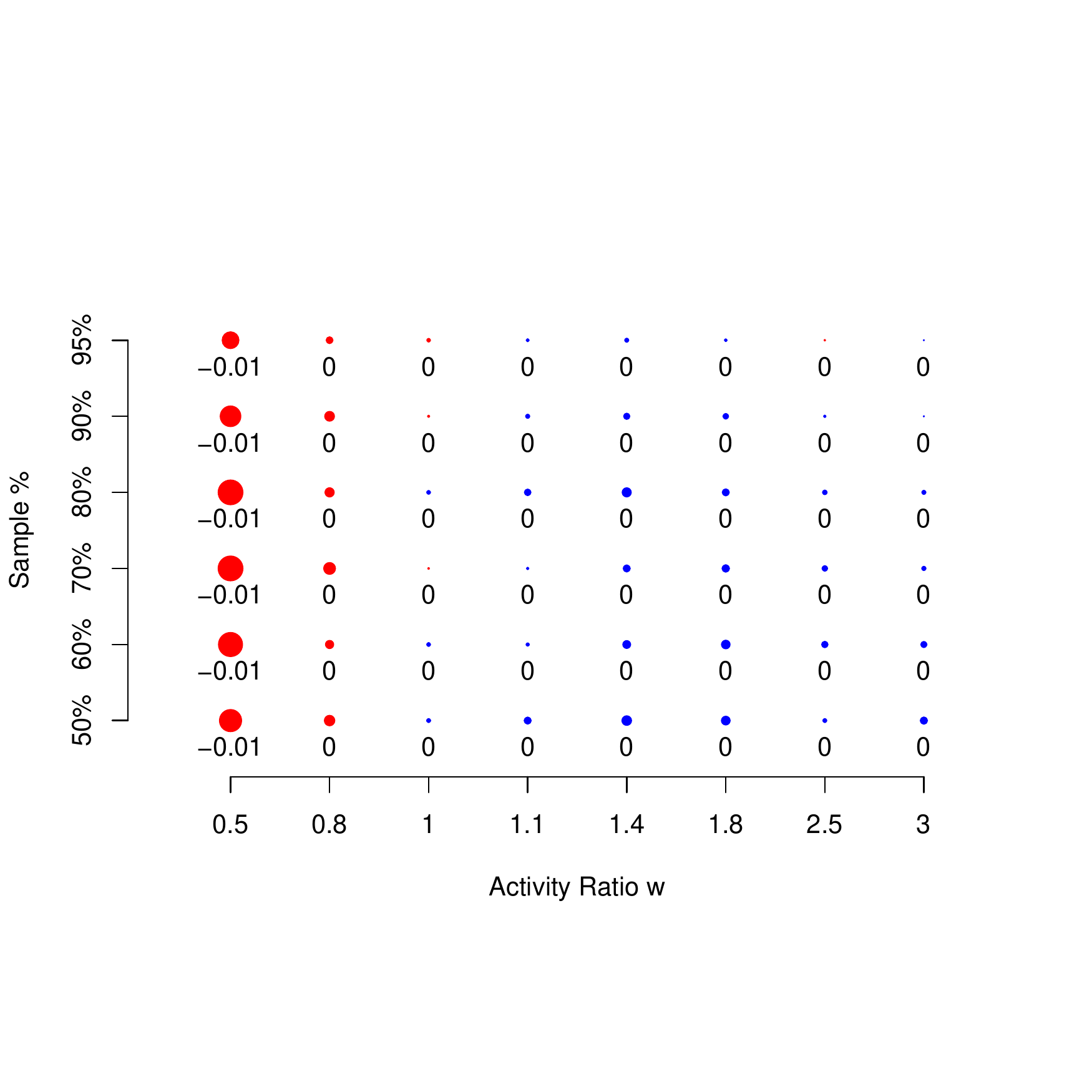}
}\end{center}  \caption{Bias of the Volz-Heckathorn and Successive Sampling estimators from samples of size 500 constituting about  50\%, 60\%, 70\%, 80\%, 90\%, and 95\% of the population, for varying activity ratio ($w$).  The same samples were used for both estimators.
}\label{bothbars}
\end{figure}

The SS estimator, however, is not subject to this type of bias, as illustrated in Figure \ref{fig:sppsbiasbars}.  This is the main contribution of the proposed estimatior.  In this plot, the bias is negligible, with the exception of the case in which $w=0.5$.  At least two factors, related to the strong homophily and the smaller size of the infected group contribute to this exception.  First, the homophily-induced dependence implies the initial sample has greater influence on the final sample than in standard successive sampling.  As this sample is selected first, its selection probabilities are closer to proportional to degree than the final successive sampling inclusion probabilities, contributing to a resulting sample slightly over-representing high-degree uninfected nodes.  Furthermore, because the infected nodes have low degrees and high homophily, they more often fail to produce both possible recruits.   
  The relative group sizes contribute to the difference in the magnitude of this effect for $w<1$ and $w>1$.

The variance of $\mh$ is also consistently lower than that of $\mvh$, combining with the lower bias to yield substantially lower mean squared error, often massively so\knote{get numbers}.\knote{put in variance plot, or some other variance measure?}  The mean squared error of $\mh$ is less than that of $\mvh$ for the full parameter space of Figure \ref{bothbars}.  The combination of bias and variance effects is visible in Figures \ref{small151} and \ref{small152}.

\newcommand{\shoe}{2in}
\newcommand{\lace}{2in}

\subsection{Sensitivity to Population Size Estimate}\label{sec:sizesens}
It is important to note that the performance of $\mh$ in Figure \ref{fig:sppsbiasbars} is dependent on knowledge of the true population size $N$.  This is often unrealistic in practice.  Therefore, we evaluate the performance of $\mh$ in the case of over and under estimation of $N$.  In particular, we consider small ($\hat{N}_s$) and large ($\hat{N}_l$) estimates of $N$ given by:
\bea
\hat{N}_s = N - \frac{N-n}{2}, ~~~
\hat{N}_l = N + \frac{N-n}{2}.
\label{nhat}
\eea

Figure \ref{small15} depicts the results of these simulations for the case of $w=1.4$.  Each sub-plot gives the distribution of the estimator over 1000 samples from each of the 6 population proportions.  The four sub-plots represent $\mvh$ (top left), $\mh(N)$ (top right), $\mh(\hat{N}_s)$ (bottom left) and $\mh(\hat{N}_l)$ (bottom right).

\newcommand{\sss}{2.1in}
\newcommand{\ttt}{2in}

\begin{figure}[h]
\begin{center}
\subfigure[$\hat{\mu}_{VH}$]
{
    \label{small151}
    \includegraphics[width=\sss,height=\ttt]{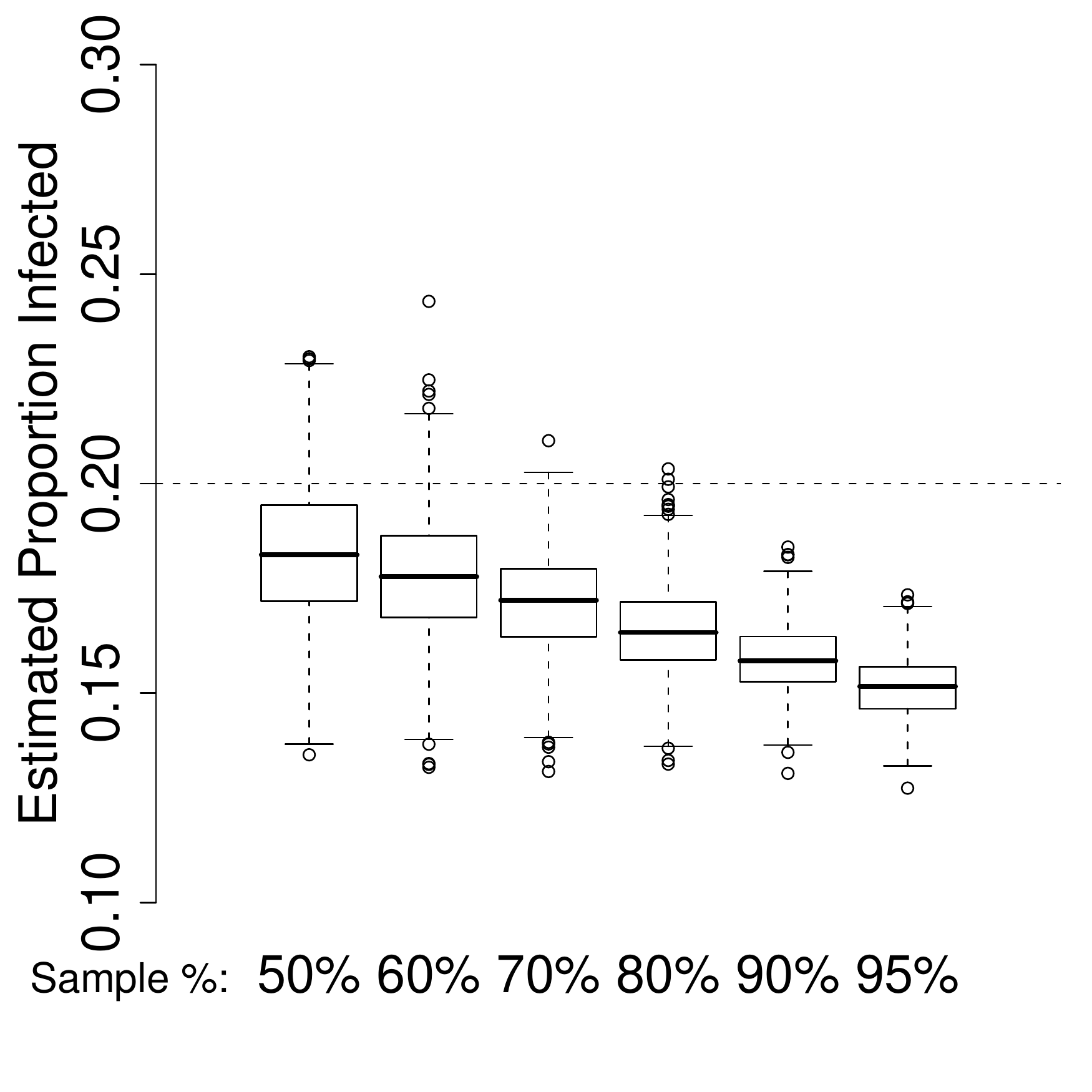}
} \hspace{.25cm}
\subfigure[$\mh$]
{
    \label{small152}
    \includegraphics[width=\sss,height=\ttt]{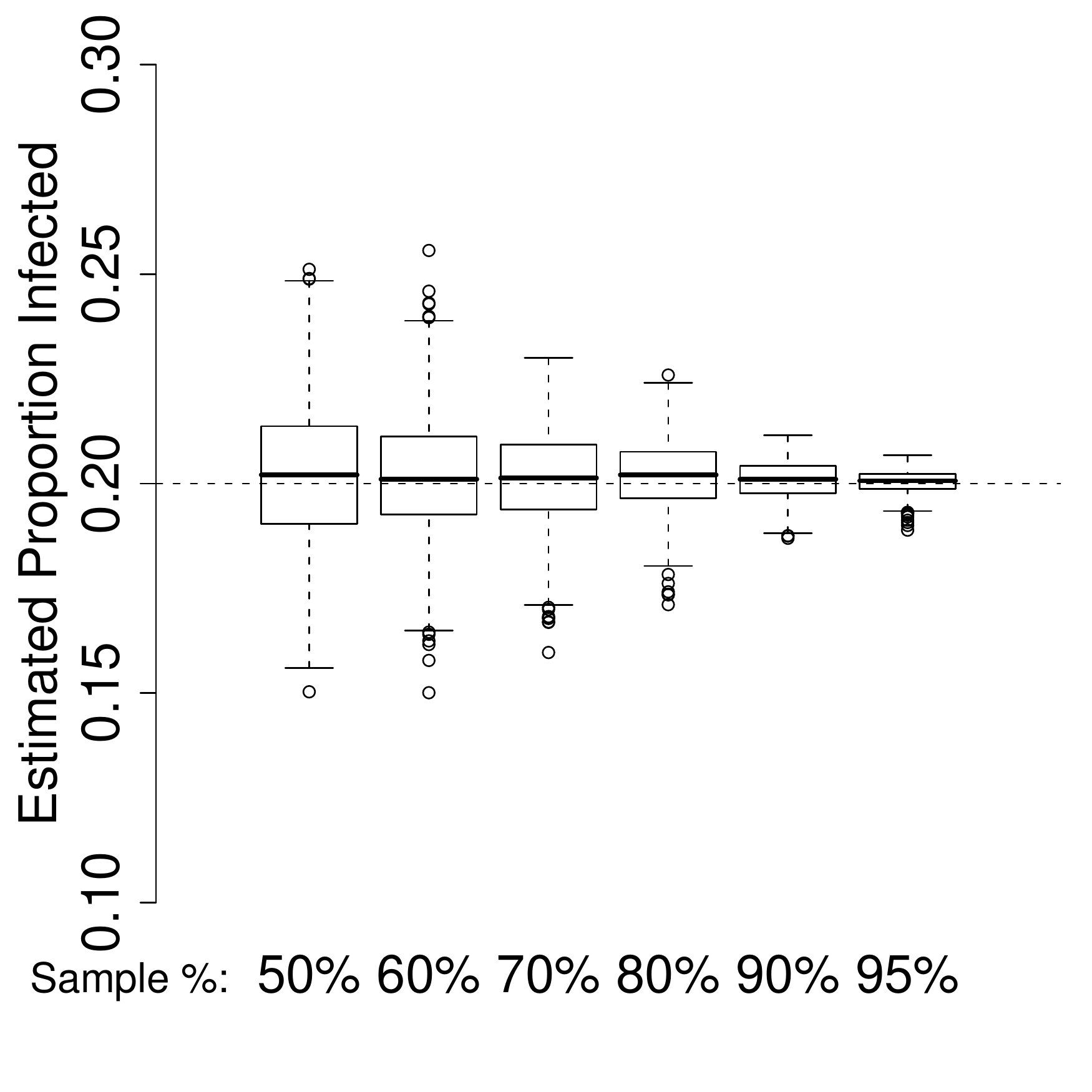}
}\vspace{.25cm}
\subfigure[$\mh(\hat{N}_s < N)$]
{
    \label{small153}
    \includegraphics[width=\sss,height=\ttt]{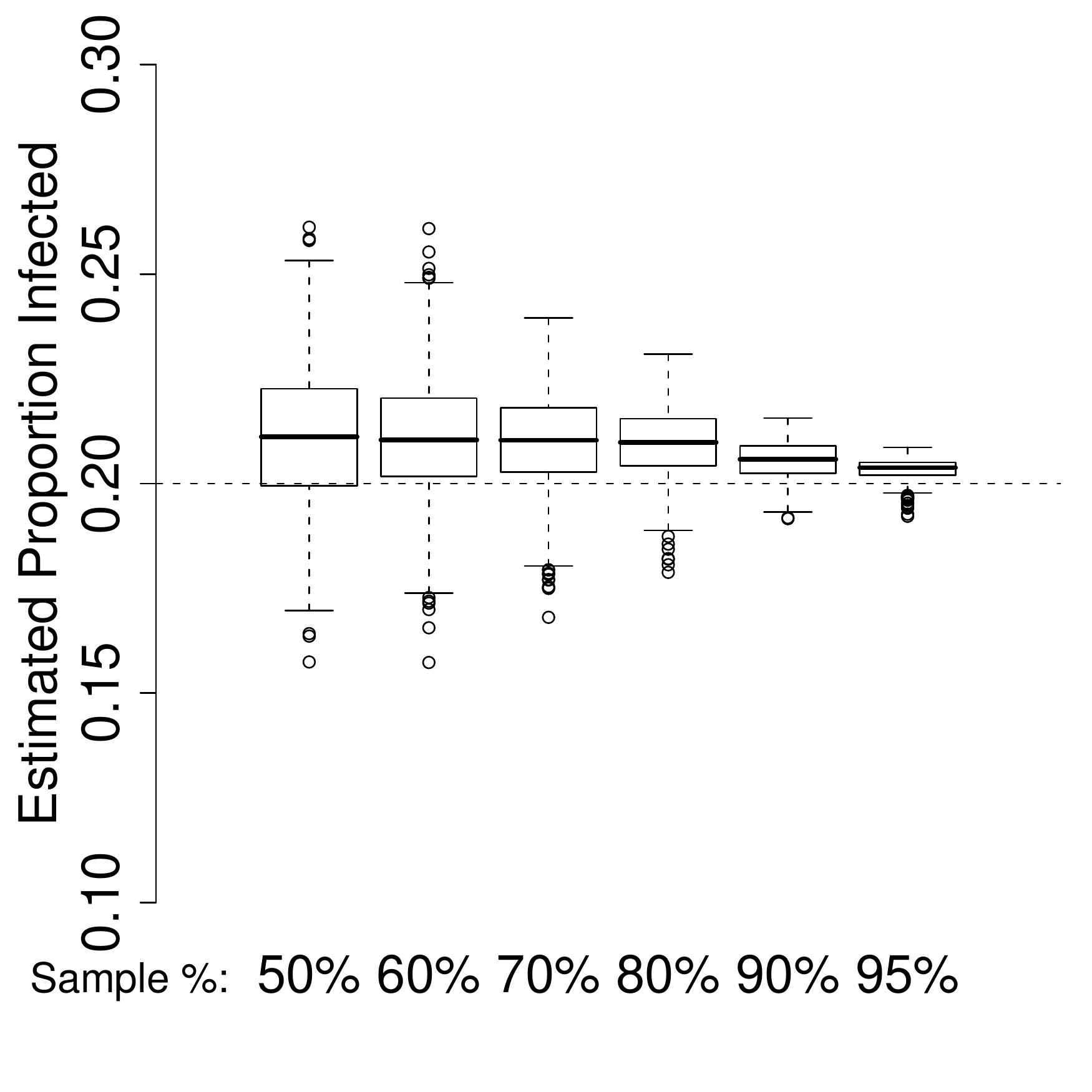}
}\hspace{.25cm}
\subfigure[$\mh(\hat{N}_l > N)$]
{
    \label{small154}
    \includegraphics[width=\sss,height=\ttt]{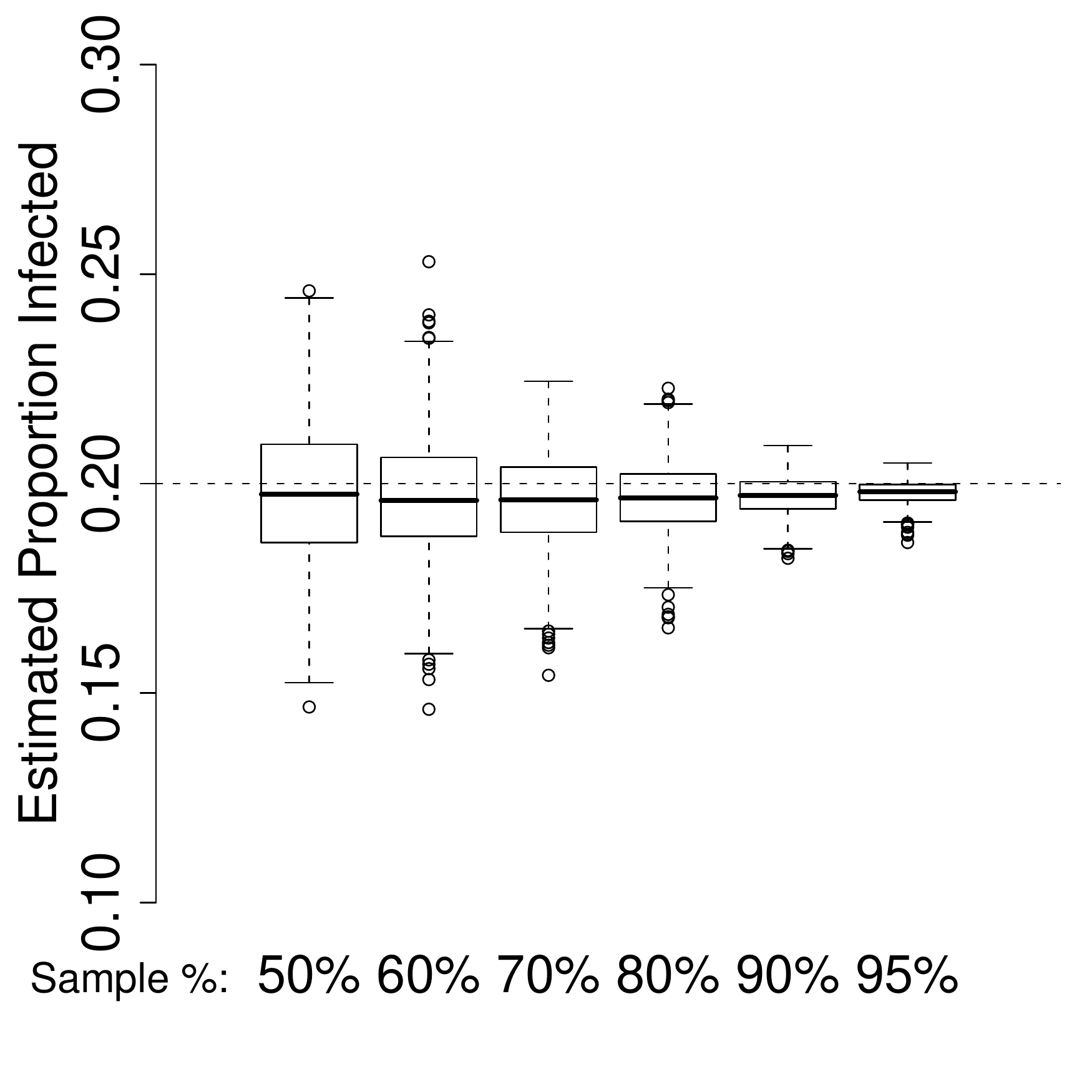}
}\end{center} \caption{$\mvh$ and $\mh$ from samples of size 500 constituting about  50\%, 60\%, 70\%, 80\%, 90\%, and 95\% of the population.  Initial samples selected independent of infection status $\vz$.  Activity ratio ($w$) fixed at 1.4. The first plots depict $\mvh$, and $\mh$.  The last two depict $\mh$ when the number of nodes is under and over estimated, respectively.  The true value is $0.20$.  The same samples were used for each estimator.
} \label{small15}
\end{figure}

In this case, the underestimation of $N$ results in a small positive bias, due to the over-estimation of the curvature of $\q$.  A small negative bias is also present in the case of $\hat{N}_l$, due to the under-estimation of the curvature or $\q$.  In both cases, the bias induced by inaccurate $\hat{N}$ is less than the bias of $\mvh$, and the resulting new estimators clearly out-perform the existing estimator.

\begin{figure}[h]
\begin{center}
    \includegraphics[width=6in]{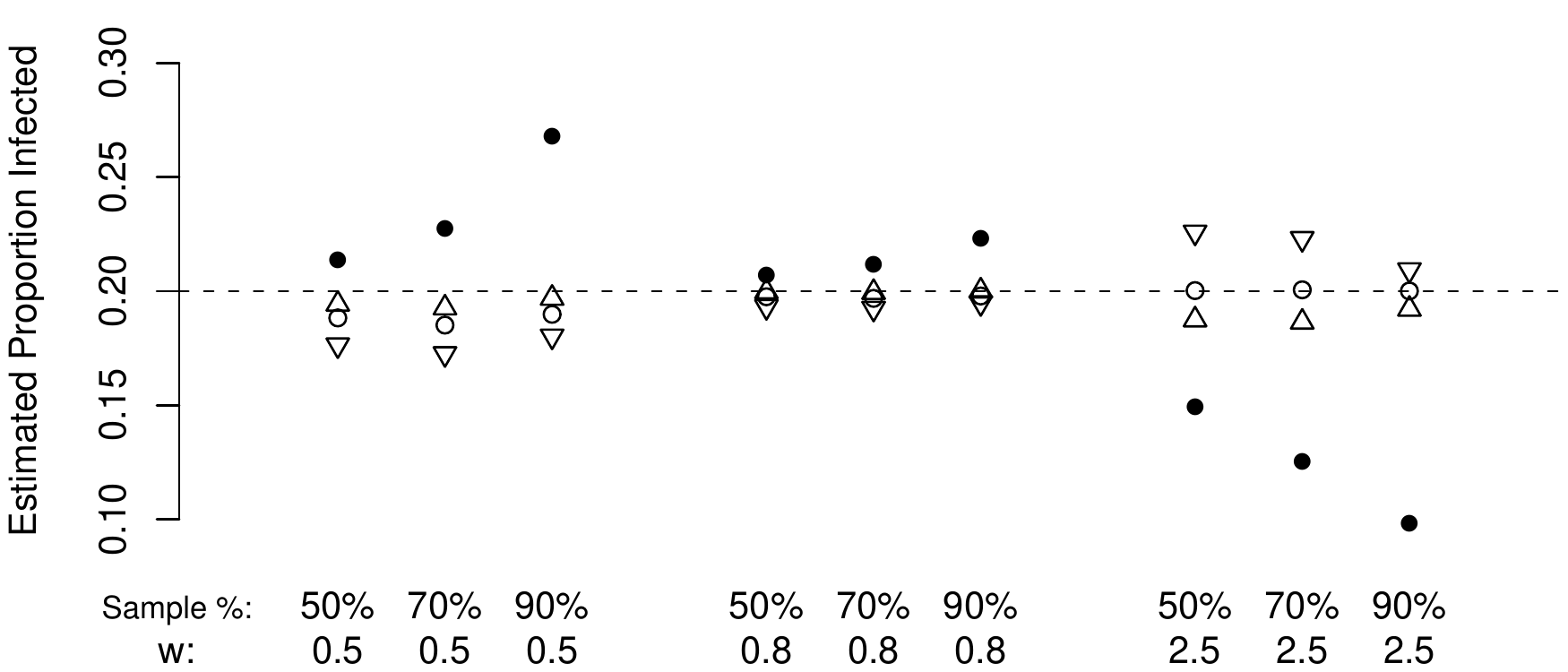}
\end{center}
\caption{Mean prevalence estimates of four types: $\mvh$ (solid circles), $\mh$ (circles), $\mh(\hat{N}_s < N)$ (down-triangles), and $\mh(\hat{N}_l > N)$ (up-triangles), considered for 50\%, 70\%, and 90\% samples, and differential activity $w = 0.5, 0.8, 2.5$.  The true value is $0.20$.} \label{fig:longguess}
\end{figure}

Figure \ref{fig:longguess} presents mean point estimates for other values of activity ratio $w$.  The middle set of plots, $w=0.8$, corresponds to a case of moderately lower mean degree among infected nodes.  In this case, the biases introduced by under and over estimation of $N$ change sign, but remain small in magnitude.  The remaining plots illustrate the greater bias possible for extreme values of $w$. While the biases become larger, they are generally smaller than those exhibited by $\mvh$, with the exception for the two cases where the negative bias of $\mh$ due to small $w$, as described in Section \ref{sec:basic} is compounded by negative bias due to $\hat{N} < N$ ($w=0.5, 0.8$, $50\%$ sample).  The mean squared error of $\mh$ is still smaller than that of $\mvh$ in these cases. Although the performance of $\mh$ is less robust to $\hat{N}$ in cases of extreme $w$, $\mvh$ also performs more poorly in these cases.

The behavior we see in these plots is typical of other simulations not shown here.  With the more extreme values of activity ratio $w$, the bias induced by the inaccurate estimation of $N$ is increased, but typically not so much as to cause performance of $\mh$ worse than that of $\mvh$.

\newcommand{\hatmu}{\hat{\mu}}
\newcommand{\mush}{\hat{\mu}_{SH}}

\knote{introduce 'seeds' or don't use it}

\subsection{Sensitivity to Initial Sample and Homophily}\label{sec:senshomoph}
The estimator $\mvh$ requires  sufficiently many sample waves to overcome the effect of an initial convenience sample and step-wise sampling probabilities not proportional to degree due to the particular network structure.
The estimator $\mh$, not based on a stationary distribution, does not allow for such an argument.  We therefore argue only that it is not worse than $\mvh$ in this respect.  In Section \ref{sec:basic}, we illustrate that in the simulated networks, bias induced by deviations from the configuration model are is no worse for $\mh$ than for $\mvh$.  We focus here on bias induced by the initial sample.
 \cite{gilehanSM09} show that $\mvh$ exhibits considerable bias in the case of a biased initial sample and network homophily.  We expect $\mh$ to exhibit similar sensitivity, and here compare its performance to that of $\mvh$.

We consider 
three regimes for the selection of the initial sample:  all uninfected, random with respect to infection, and all infected.  We also consider five levels of homophily, measured by the ratio $R$ defined by:
\bea
R = \frac{\textrm{Probability of an {\ql}infected-infected\qr tie}}{\textrm{Probability of an {\ql}infected-uninfected\qr tie}}, ~~~~ R=1,2,3,5,13.
\eea
The standard level of homophily used in this paper corresponds to $R=5$.
  To present a comparison most favorable to $\mvh$, we treat a population size of 1000, (a $50\%$ sample), and activity ratio $w=1$.

We consider 1000 samples from each homophily and sampling scenario, and summarize performance in terms of absolute bias, variance, and mean squared error, the last of which is depicted in Table \ref{tab:mse2}.  We find that the variance of $\mvh$ always exceeds that of $\mh$, as does bias in all but a few cases of small differences. The MSE of $\mvh$ always exceeds that of $\mh$.  Table \ref{tab:mse2} illustrates this relation for the case of an all-infected initial sample.  The patterns for the other two sampling conditions are similar.

\begin{table}[h]\caption{Comparison of mean squared error for $\mvh$ and $\mh$ with  all infected initial sample and five levels of homophily. $MSE(\mh)<MSE(\mvh)$ for all conditions.}
\begin{center}
\begin{tabular}{l||ccccc}
Homophily ($R$) & 1 & 2 & 3 & 5 & 13 \\
  \hline
  \hline
MSE $\mvh$ & 0.00029 & 0.00036 & 0.00054 & 0.00150 & 0.01525 \\
  MSE $\mh$ & 0.00026 & 0.00032 & 0.00052 & 0.00140 & 0.01449 \\
  \hline
  Efficiency: MSE $\mvh$/MSE $\mh$ & 1.12 & 1.12 & 1.04 & 1.07 & 1.05 \\
\end{tabular}
\end{center}\label{tab:mse2}
\end{table}
\knote{for now ignore that S-H bias in the opposite direction}


\newcommand{\s}{\mathscr{S}}
\newcommand{\qys}{\q^*_{k,z}(y,\s)}
\newcommand{\qysk}{\q^*_{k,z}(y^i,\s^i)}

\section{Application to HIV Prevalence in High-Risk \\Populations}\label{sec:apply}

\subsection{Background}
The United Nations requires countries to measure and monitor key indicators related to their HIV epidemics \citep{UNAIDS2007, UNAIDS2008a}.  In particular, countries with epidemics concentrated in high-risk groups are required to report on several features of key populations such as (injecting) drug users ((I)DU), men who have sex with men (MSM), and sex workers (SW).  Such features include HIV prevalence, risk behaviors, and population sizes.  Because these populations are typically hard-to-reach, many countries rely on respondent-driven sampling to estimate HIV prevalence and risk behaviors.  In this section, we evaluate data collected on IDU and MSM in cities from two different countries in 2007 and 2008.

\begin{figure}[h]
\begin{center}
\subfigure[East Europe IDU]
{
    \label{fig:eeidu}
    \includegraphics[width=7.5cm]{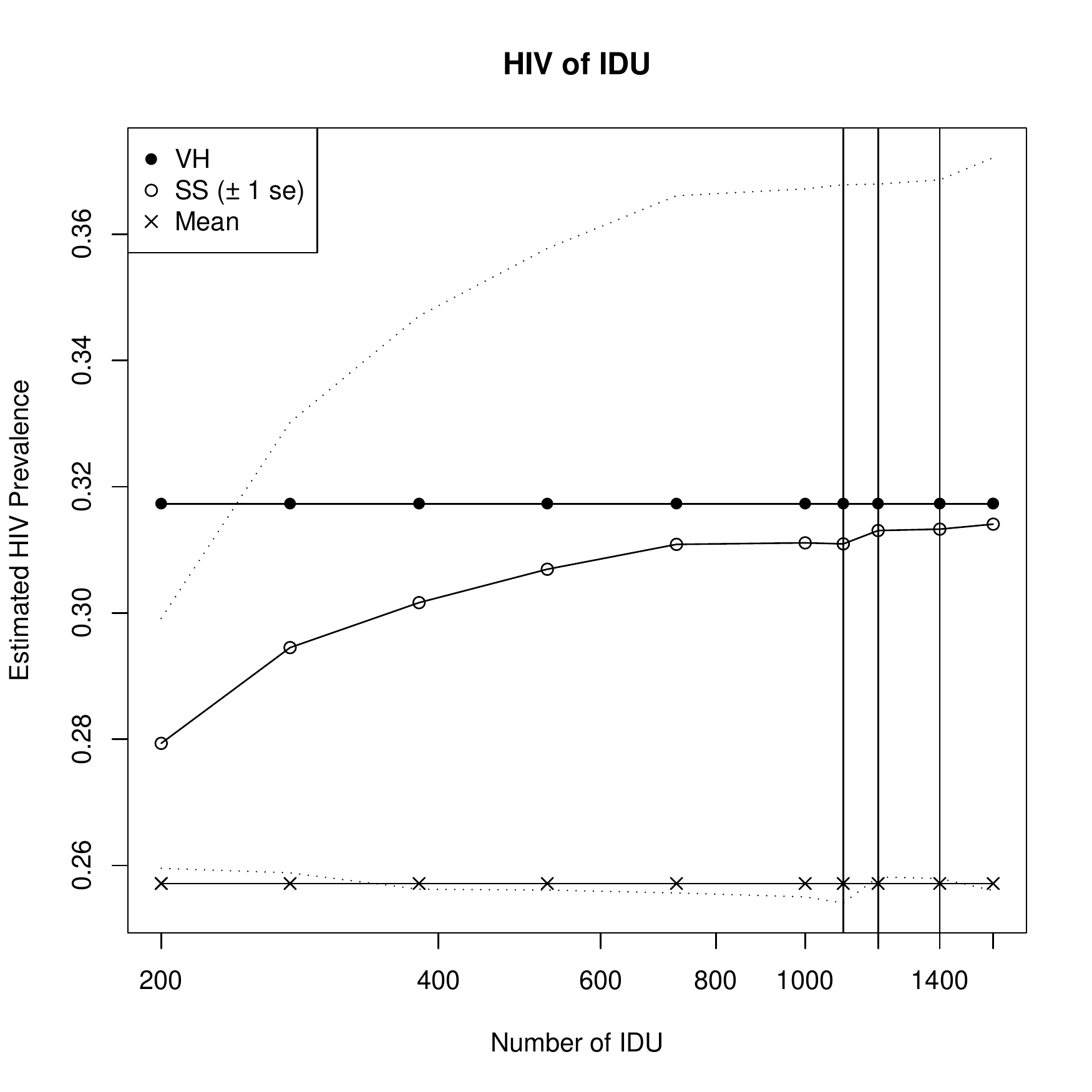}
} \hspace{-.5cm}
\subfigure[Caribbean DU]
{
    \label{fig:cdu}
    \includegraphics[width=7.5cm]{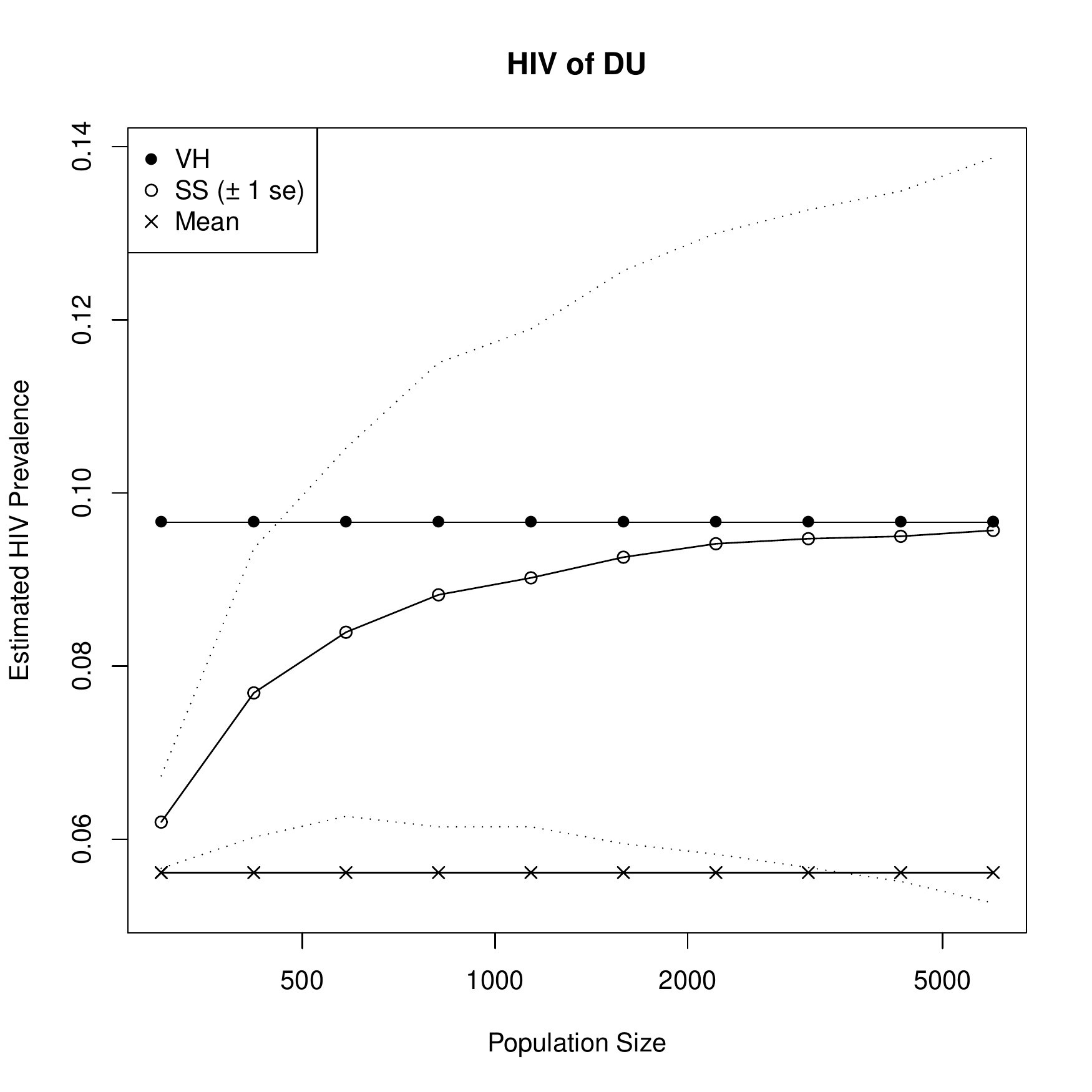}
}\hspace{-.5cm}
\subfigure[Caribbean MSM]
{
    \label{fig:cmsm}
    \includegraphics[width=7.5cm]{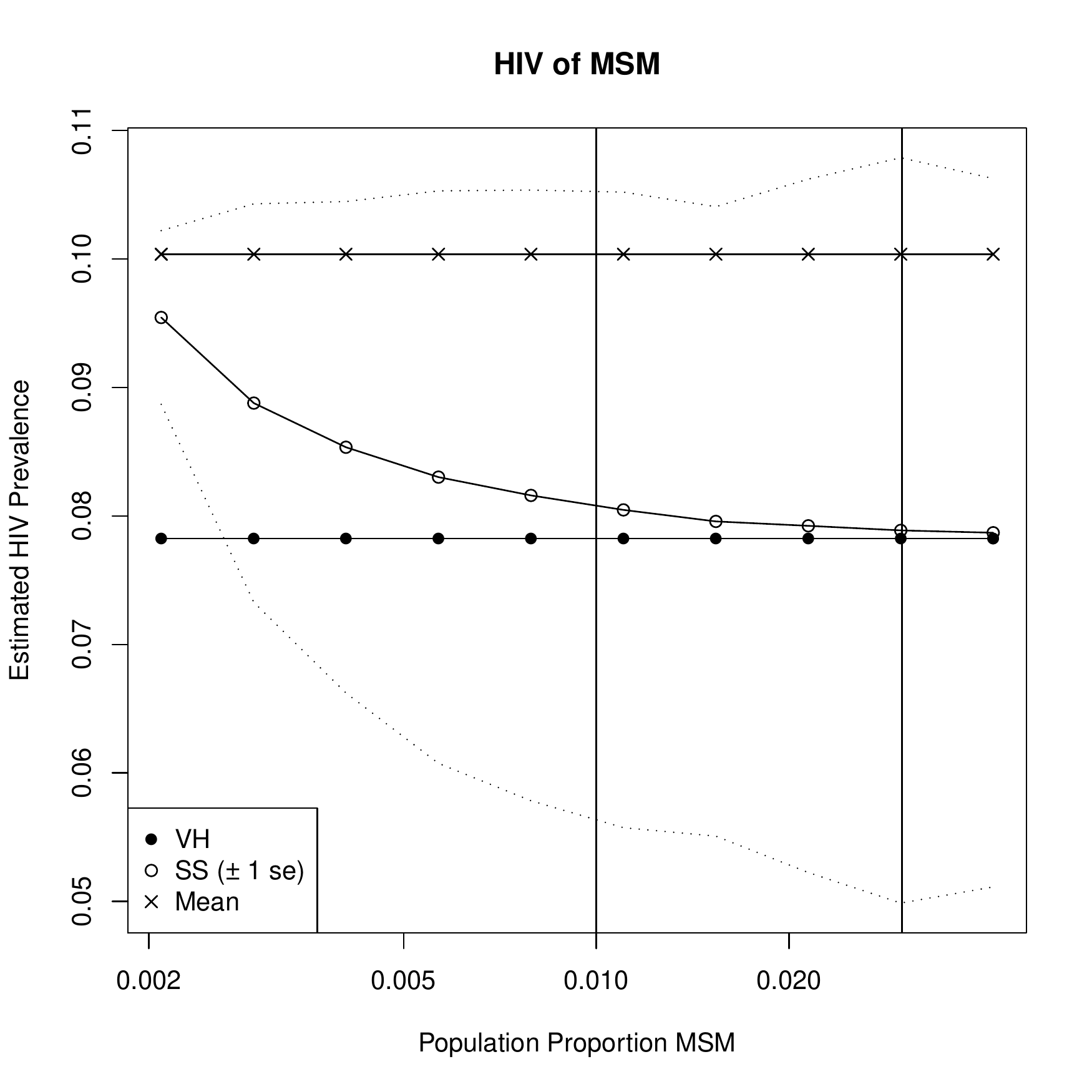}
}\end{center}  \caption{Estimated HIV prevalence in three populations according to $\mh$, $\mvh$, and the sample mean, for various population size estimates.  Dotted lines represent 1 standard error above and below $\mh$, according to the estimator in the supplemental materials.  Vertical bars on (a) represent exogenously estimated population size and on (c) represent rule-of-thumb 1\% to 3\% MSM in the population.
}\label{fig:examples}
\end{figure}

\subsection{Injecting Drug Users in an East European City}
The first example is based on a 2007 survey of IDU in major cities in a former Soviet block country.  The HIV epidemic in this country is largely driven by IDU, where prevalence levels are over 50\% in many cities, and RDS is used to study this population, as well as MSM and SW in several larger cities.  This country also invests heavily in producing estimates of the size of the hidden population through scale-up and multiplier methods \citep{Kruglov08, UNAIDS2003}.  We focus here on the results for one city.  Because in very large populations we expect $\mh$ to be nearly identical to $\mvh$, we consider a city with one of the smaller estimates for the size of the hidden population.  In this city, the number of IDU is estimated at 1200, with confidence interval 1100-1400.  The RDS sample began with 6 initial samples, and distributed 3 coupons per respondent.  The sample size was 175, all with full degree and HIV data.  The two longest sample chains ended at wave 9, with 4 total respondents from wave 9.

We estimate the HIV prevalence using $\mh$ for several population sizes including 1100, 1200, and 1400, and also report the corresponding estimates of $\mvh$ and the sample mean.  These results are summarized in Figure \ref{fig:eeidu}.  We find that the prevalence estimate based on $\mh$ is very close to that based on $\mvh$ for population sizes in this range, and well within the uncertainty of the estimate, according to the estimator in the supplemental materials.  This pattern is consistent across the cities we have considered in this country.  This is partly due to the policy in this country of using RDS only in cities with larger estimated population sizes, opting for strategies of institutional sampling or attempted complete enumeration in areas with smaller populations.


\subsection{Drug Users in a Caribbean City}
The second example is taken from a small Caribbean country, which has conducted RDS studies of drug users (DU), MSM, and SW in four main cities in 2008 (note that this study did not limit participation to {\it injecting} drug users).  We focus here on the study of DU in one of the cities with smaller total population.  In this study, there were 7 initial samples, resulting in a sample of size 301, of which we include here only the 285 with full degree and HIV information.  Again three coupons were distributed to most respondents, although this number was reduced as the sample size approached the target of 300.  The two longest sample chains reached wave 11, with 6 respondents from this wave.

In this city, the number of DU is unknown.  Therefore, we use the successive sampling estimator to produce a sensitivity analysis for the effect of population size on the HIV prevalence estimate.  The results of this analysis are shown in Figure \ref{fig:cdu}.
Here the point estimate varies from 6.2\% for a population size of 301 to 9.6\% for a population size of 6000, with $\mvh = 9.7\%$.  In absolute and relative terms, this is more variation than in the previous example, (3.4\% or about 15\% of $\mvh$).  Although these differences are still well within the uncertainty of the estimator,  in many cases RDS point estimates are important in themselves.  It is also possible that a sensitivity analysis such as this one will be of interest with respect to a particular prevalence threshold, such at the $5\%$ threshold on which UNAIDS bases national epidemic classifications.  In this example, the point estimate of prevalence is above $5\%$ for all population sizes, while a nominal 90\% confidence interval includes 5\% for all population sizes.

\subsection{Men who have Sex with Men in a Caribbean City}
Finally, we consider a population of MSM in the same Caribbean city.  This study design was very similar to the corresponding DU study, with 7 initial samples, and a maximum of 3 coupons distributed.  Here, the sample size was 270, with complete information for 269.  One chain in this sample reached wave 12, with two respondents in that wave.

Figure \ref{fig:cmsm} illustrates the sensitivity of the prevalence estimate to the number of MSM in this city.  For comparison, the horizontal axis corresponds to the same population sizes as in \ref{fig:cdu}, although here it is labeled in proportions. The prevalence estimate changes by about 1.6 percentage points over this range of possible population sizes (9.5\% for 301 or .2\% of the population are MSM, to 7.9\% for 6000 or 4.2\% of the population are MSM) .  There are no additional studies available to provide population size estimates, however this problem is less severe for populations of MSM, as a rule-of-thumb is often applied, estimating the number of MSM at about 1\%-3\% of the general population, corresponding to the two vertical bars on Figure \ref{fig:cmsm}.  In this case, however, further information is available.  This sample did not reach its desired sample size (300), because no more coupons were returned, and the research team was not able to find additional MSM to sample.  Thus the sample neared exhaustion of the portion of the population available for sampling, suggesting that although the population of MSM might be 1-3\% of the city population, a smaller portion of those may be connected to the giant component of the social network of MSM and willing to participate in such a study.  Therefore, if we restrict our inference to the reachable portion of the target population, the relevant population size is likely much closer to the sample size, suggesting a value of the $\mh$ estimate over 1.5 percentage ponts higher than $\mvh = 7.8\%$.
Note that among the 12 RDS studies in this country, three samples exhibited this near exhaustion behavior.  

It is also of interest to note that in this example, unlike the previous two, $\mh$ is consistently higher than $\mvh$.  This result is consistent with the expectation that $\mh$ is typically between $\mvh$ and the sample mean.  In this case, the degree-based weights of $\mvh$ reduced the overall weight given to infected nodes, and the more moderate weights used in $\mh$ reduced this effect.

\section{Discussion}\label{sec:discussion}
The key insight of this paper is the recognition that the true mapping $\q$ from nodal degree $\vd_i$ to inclusion probability, $\vpi_i$ under Respondent-Driven Sampling is better approximated by successive sampling than by the linear mapping assumed by \cite{volzheck08}.  In addition, we introduce a novel approach to estimating the unit size (or degree) distribution in the population, based on the population size and sizes of observed units under successive sampling.  Combining this insight and estimation strategy, we present a new estimator for population proportions based on an RDS sample.

\begin{figure}[h]
\begin{center}
    \includegraphics[width=6in]{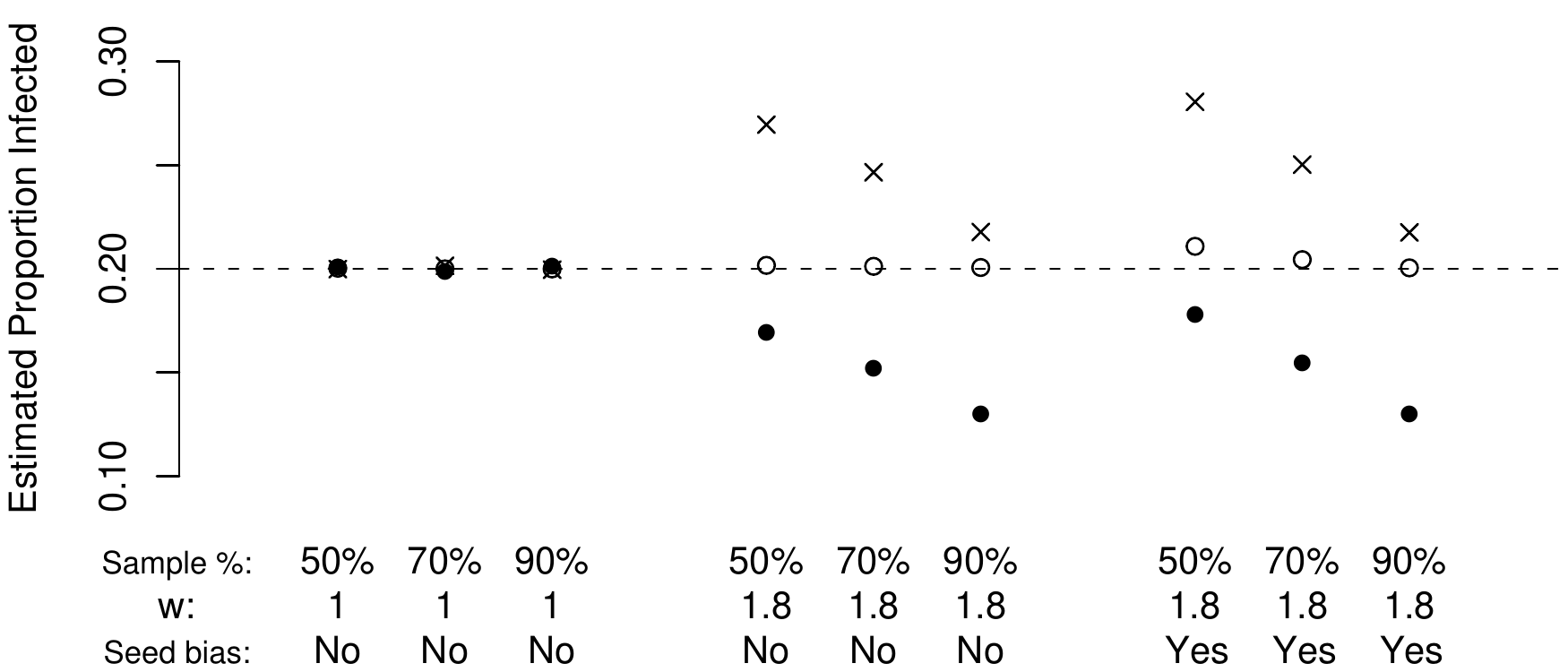}
\end{center}
\caption{Mean prevalence estimates of three types: $\mvh$ (solid circles), $\mh$ (circles), and sample mean $\hat{\mu}$ (crosses), considered for 50\%, 70\%, and 90\% samples, and activity ratio $w = 1, 1.8$, and for initial sample (seeds) selected at random with respect to infection and all infected.  None of the estimators exhibit bias in the face of homophily with $w=1$.  When $w \neq 1$, $\mvh$ and $\hat{\mu}$ exhibit bias, with magnitude sensitive to sample fraction.  All estimators exhibit bias in the case of biased initial sample.} \label{fig:long}
\end{figure}

The contribution of this new estimator is illustrated in Figure \ref{fig:long}.  In cases with no correlation between degree and quantity of interest ($w=0$), and initial sample selected at random with respect to infection status, the naive sample mean, $\hat{\mu}$ is an unbiased estimator of the population mean, as are existing estimators such as $\mvh$ and the proposed estimator $\mh$.  This is illustrated in the first three columns of the figure.

However, when the variable of interest is related to nodal degree, classes with higher degrees are over-represented in the population mean, resulting in the positive bias in the sample mean $\hat{\mu}$ in the second three columns of Figure \ref{fig:long}.  This bias shrinks as the sample fraction increases.  The existing estimator $\mvh$ adjusts reasonably well for this effect when the sample fraction is small, however for larger sample fractions, this estimator over-compensates for the effect of degree distribution, resulting in bias opposite that of $\hat{\mu}$. The contribution of the proposed estimator, $\mh$ is that it correctly adjusts for the joint effects of varying degree distributions and large sample fractions.

The last three columns of Figure \ref{fig:long} illustrate a shortcoming of all three of these estimators.  In this case, all initial samples, or {\it seeds}, are selected from among the infected nodes.   This results in increased positive bias in all three of these estimators.  All three estimators are also subject to other sources of bias discussed in \cite{gilehanSM09}, including bias induced by the systematically biased passing of RDS coupons.

Because RDS is in wide usage, and often in cases where the sample fraction is large, this new estimator may improve estimation in many contexts.  Its applicability is limited, however, by the requirement that the population size $N$ is known.  Figures \ref{small15} and \ref{fig:longguess} illustrate that inaccurate estimates of $N$ can introduce bias into $\mh$, although this new estimator still out-performs $\mvh$ with the level of inaccuracy considered in this study.  

While most of our simulation study has focused on sample fractions consistent with the finite population effects that $\mh$ is intended to address, it is of interest to note that $\mh$ is nearly identical to $\mvh$ for small sample fractions.  We have replicated much of the simulation study here with population size $N=10,000$.  We find no significant differences between $\mh$ and $\mvh$ in these simulations, over a full range of values of $w$.  When using inaccurate estimates of $N$ as in (\ref{nhat}), we found only one case of significant difference between $\mvh$ and $\mh$, corresponding to a dramatic under-estimate of population size, $\hat{N_s}=5250, N=10,000$, and for high activity ratio, $w=3$.  In this case, $\mh$ had bias $0.0139$ and $\mvh$ had bias $0.0105$.  The difference in mean squared error was not significant.  Thus, the proposed estimator is helpful in correcting for finite population biases when they exist, and sensitive to the estimated population size in these cases.  In cases where finite population effects are not present, the proposed estimator nearly coincides with existing estimator $\mvh$, and, unless the inaccurate population size estimate is small enough to transition to the region of finite population sensitivity, $\mh$ is insensitive to population size estimates in these cases.

Known population size $N$ and random mixing are key assumptions of the estimator $\mh$.  Additional assumptions required by $\mh$ are listed in Table \ref{tab:assmh}. 
The assumptions with \sout{strikethrough} are necessary for $\mvh$ but not for $\mh$, and the 
{\it italic text} indicates 
additional assumptions required by $\mh$ but not $\mvh$. In these terms, the key contribution of $\mh$ is to remove the dependence on a known inaccurate random walk model for estimating sampling probabilities.  In return, $\mh$ sacrifices the theoretical robustness to the initial sample promised by the Markov chain model, although this robustness was not truly present in $\mvh$ either, 
and the proposed estimator performs no worse than the former in this respect (Table \ref{tab:mse2}).  More critically, $\mh$ relies on the assumption of known population size.  The estimator $\mh$ is also sensitive to the degree of non-random mixing, or homophily in the network, although no more sensitive than $\mvh$ (Table \ref{tab:mse2}), as well as to other assumptions such as accurate self-reported degree and an undirected network.  \cite{gilehanSM09} illustrate the sensitivity of $\mvh$ to deviations from some other assumptions in Table \ref{tab:assmh}.  We do not expect $\mh$ to be more robust to such deviations than $\mvh$.

\newcommand{\etwo}{}

\newcommand{\ecol}{}
\newcommand{\eone}{}

\begin{table}\caption{Assumptions of $\mh$. Assumptions with \sout{strikethrough} apply to $\mvh$ but not $\mh$.  Assumptions in the {\it italic} apply to $\mh$ but not $\mvh$.}
\begin{center}
\begin{tabular}{l||c|c}
& Network Structure  & Sampling Assumptions\\
& Assumptions & \\
\hline
\hline
Random Walk & \ecol \eone \sout{Population size large ($N >> n$)} & \ecol \eone \sout{Sampling with replacement} \\
Model & 
& \eone \sout{Single non-branching chain} \\
\hline
Remove Initial & \ecol \etwo Homophily weak enough & \ecol \etwo Sufficiently many sample waves \\
Sample & Connected graph & \it  Initial sample unbiased\\
Dependence & & \\
\hline
To Estimate   & All ties reciprocated & Degree accurately measured  \\
Probabilities & \it Known population size $N$ & \ecol \etwo Random referral   \\
\end{tabular} \label{tab:assmh}
\end{center}
\end{table}

\renewcommand{\etwo}{\gray}

Our application to HIV prevalence estimation in Section \ref{sec:apply} illustrates several uses for $\mh$.  First, when the population size is known, $\mh$ provides a better estimate of population proportions than the other available methods.  This can also be done when only a range of population sizes is known.  In particular, in the case of MSM, officials are often willing to assume that the population ranges between 1\% and 3\%.  When no information is available on population size, $\mh$ can be used to perform a sensitivity analysis.  Finally, other information from the sampling process, such as the exhaustion of the population available for sampling, may suggest that the sample fraction is large.  
Additional information may also be gathered by asking respondents about their exposure to others who have been sampled.

Intuition suggests that in the case of a large sample fraction, an RDS estimator should negotiate between the infinite-population assumption of the Volz-Heckathorn estimator and the full-population sample assumption of the naive sample mean.  In this paper, we provide an estimator based on a successive sampling model that appropriately negotiates between these two extremes.  Whenever the hidden population size is known, it will be preferable to use this estimator.  When the hidden population size is not known, this estimator still provides a helpful diagnostic check on the other available estimators.  

Beyond the estimator itself, this paper contributes a new theoretical framework for understanding the sampling process in respondent-driven sampling.  Previous understandings have relied on a with-replacement or infinite population assumption, an assumption known to be inaccurate, and critically so in some populations.  The introduction of a sampling model appropriately accommodating the true without-replacement nature of the sampling process opens new possibilities for future research on respondent-driven sampling.  We intend to make code available for these procedures in the R package {\tt RDS} on CRAN.

\singlespacing
\bibliography{../networks}

\end{document}